\documentclass{ws-procs9x6-mod}

\begin{document}

\title{
 Integration of Acoustic Detection Equipment into ANTARES
\footnote{
\uppercase{T}his work was supported by the \uppercase{G}erman \uppercase{BMBF}
\uppercase{G}rant \uppercase{N}o. 05~\uppercase{CN}2\uppercase{WE}1/2.
}
}

\author{R.~Lahmann, G.~Anton, K.~Graf, J.~H{\"o}ssl, A.~Kappes, T.~Karg, 
U.~Katz, C.~Naumann and K.~Salomon}

\address{Physikalisches Institut,\\ Friedrich-Alexander-Universit{\"a}t 
Erlangen-N{\"u}rnberg,\\
 Erwin-Rommel-Stra{\ss}e 1,\\ 91058 Erlangen, Germany\\ 
E-mail: Robert.Lahmann@physik.uni-erlangen.de}

\maketitle

\abstracts{ 
The ANTARES group at the University of Erlangen is working
towards the integration of a set of acoustic sensors into the ANTARES
Neutrino Telescope\cite{ANTARES}. With this setup, tests of acoustic
particle detection methods and background studies shall be performed.
The ANTARES Neutrino Telescope, which is currently being constructed
in the Mediterranean Sea, will be equipped with the infrastructure to
accommodate a 3-dimensional array of photomultipliers for the
detection of Cherenkov light. Within this infrastructure, the required
resources for acoustic sensors are available: Bandwidth for the
transmission of the acoustic data to the shore, electrical power for
the off-shore electronics and physical space to install the acoustic
sensors and to route the connecting cables (transmitting signals and
power) into the electronics containers.  It will be explained how the
integration will be performed with minimal modifications of the
existing ANTARES design and which setup is foreseen for the
acquisition of the acoustic data.  
}

\section{Introduction}

A promising alternative to neutrino telescopes detecting Cherenkov
light in a transparent medium (ice, fresh water, sea water) with
optical attenuation lengths of several tens of meters arises from the
fact that particle showers with energies exceeding values in the order
of 100\,PeV produce detectable bipolar pressure waves in water of
about $50\,\mu$s length with a range of up to several km.  This effect
is described by the thermo-acoustic model and has been experimentally
verified for proton and laser beams\cite{Graf}.  Acoustic neutrino
telescopes therefore might allow for future giant-volume detectors in
the order of 1000\,km$^3$.

A necessary prerequisite to develop acoustic detection methods is the
detailed understanding of background conditions and the investigation
of signal identification methods.  In order to acquire the long-term,
high-precision data needed for this purpose, it is intended to
instrument a part of the ANTARES detector with acoustic sensors.

\section{The ANTARES detector and its data acquisition system}

The ANTARES detector is currently under construction in the
Mediterranean Sea, off-shore of Toulon, and is connected to the coast
by an electro-optical cable of about 40\,km length.  The instrumented
area will range from a depth of about 2000\,m to 2400\,m.  The
detector is designed to detect the Cherenkov light from muon tracks.
For this purpose, it will be equipped with a total of 900 optical
modules\cite{Amram:2001mi} (OMs), which hold one photomultiplier tube
(PMT) each inside a sphere, pointing downwards at an angle of
45$^\circ$.

\begin{figure}[ht]
\centerline{
\includegraphics[width=4.4in]{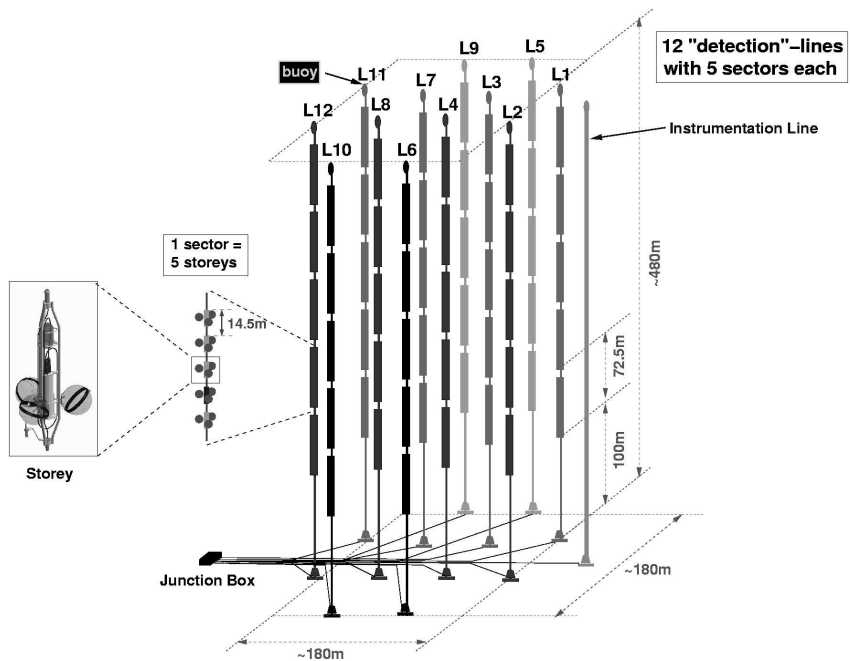}
}
\caption{
Schematic view of the ANTARES detector.  Shown are the 5 sectors per
detection line and the 5 storeys per sector.  The second storey from
the bottom of each sector contains the Master LCM (MLCM) for data
transmission to shore (see text).  Also shown is one storey with its 3
optical modules (OMs) that contain one PMT each (not shown).  The
connection to the shore from the Junction Box is established via a
$\sim$40\,km cable comprising optical fibres and electric power lines
(not shown).  Equipment on the Instrumentation Line is not shown.
\label{fig-ANTARES}
}
\end{figure}

When finished, the detector will consist of 12 ``detection lines'',
arranged in an octagonal shape on the seabed at a distance of about
70\,m from each other (cf. Fig~\ref{fig-ANTARES}).  In addition, one
``instrumentation line'' will hold equipment to record environmental
conditions such as the current profile and the salinity of the sea
water.

Each detection line comprises 25 storeys at a distance of 14.5\,m from
each other. Each storey consists of a mechanical support structure
that holds 3 OMs and a titanium container with the required
electronics (``local control module'', LCM).
Five storeys form a sector which constitutes one unit for
purposes of data readout.
It is foreseen to integrate the acoustic detection equipment into
ANTARES in the form of ``acoustic sectors'' with acoustic sensors
replacing the PMTs and using as much of the infrastructure provided by
ANTARES with as little changes as possible.  This will be described in
detail below.

Each LCM contains a backplane that is equipped with connectors for the
electronics cards and provides power and data lines to and from the
connectors.  In each sector, one LCM is designed as Master LCM (MLCM)
which in addition to its data taking tasks manages the data
transmission to the shore through single-mode optical fibres using
TCP/IP.
The main electronics cards of a LCM are:
\begin{itemize}
\item
A clock card, which provides a time stamp to each recorded PMT-value
with a resolution of 50\,ns and a precision of about
0.5\,ns\cite{Blanc:2003mv};
\item
So-called ARS-boards which contain 2 Analogue Ring Sampler (ARS)
ASICs\cite{Druillole:2001dm} each, conditioning and digitising the
analogue data from the PMTs. Furthermore, they subdivide the 50\,ns
clock signal into 256 steps.  In each LCM, 3 ARS boards are installed
to read out the 3 PMTs per storey.
\item
A data acquisition or DAQ card, which reads out the ARS boards and
provides the communication to the MLCM of the sector via TCP/IP.
\end{itemize}

In the standard sampling mode, an ARS-chip is triggered by a PMT
signal above a predefined threshold and then employs a pulse shape
discriminator to recognise single photon events---for which the
integrated analogue charge is then digitised and read out.  The data
rate, which is dominated by background from $^{40}$K and
bioluminescence, therefore can be adjusted by varying the threshold
for single photon events.

The bandwidth of the complete data acquisition chain is limited by the
throughput of the DAQ-boards which---as will be explained
below---limits the number of acoustic sensors that can be installed
per storey.

\section{Integration of acoustic sensors and their readout}

The acoustic sensors that are currently developed and tested by the
ANTARES group at Erlangen are described in Ref. \refcite{Naumann}.
The two design concepts (individual hydrophones and ``acoustic
modules'', in which acoustic sensor elements are installed in the
spheres of the optical modules instead of the PMTs) do not differ in
their requirements for the electronics inside the LCM.

The fundamental guideline for the design of the acoustic detection
system has been that the implementation shall be done with as few
modifications to the existing ANTARES design as possible.  These
considerations have lead to the following layout principles:

\begin{itemize}
\item
In order not to compromise the suitability for a deep-sea environment,
no modifications must be done to the titanium container or the
penetrators and connectors leading into and out of the container.
Consequentially, only the 3 holes that are present in each container
for the cables leading to the 3 OMs can be used to connect to the
acoustic sensors.
\item
In the LCMs of the acoustic storeys, the ARS boards will be replaced
by ``Acoustic ADC boards''; no other changes to the electronics will
be done.  These boards will digitise the acoustic data and format
them, where the format will be exactly the same as that of the optical
data from the ARS boards.
The acoustic data will then be read out sequentially by the DAQ board
and transmitted to shore in exactly the same fashion as the PMT data.
\item
On-shore, the separation of acoustic and optical data will be based on
their origin, i.e. on the IP address of the DAQ-board in the
corresponding LCM.  Acoustic data will be separated from the main data
stream and processed, filtered and compressed on a dedicated PC-farm.
\end{itemize}

The number of hydrophones per storey is limited by the data throughput
of the DAQ-board processor of roughly 20-25 Mb/s. Consequentially, 6
acoustic sensors per storey can be installed for a 16-bit digitisation
and a 200\,kHz sampling rate.  The sampling rate can be further
increased if an adequate down-sampling is performed on the acoustic
ADC board.

Each of the 3 acoustic ADC boards per LCM will contain two 16-bit ADCs
with a maximum sampling frequency of 500\,kHz for the processing of
two acoustic sensors.  The data sampling will not be triggered but
instead be continuous at an adjustable data rate of 100\,kHz, 200\,kHz
or 400\,kHz.  It will be possible to individually disable each
hydrophone.

In order to minimise the development time and error-proneness while maximising 
the flexibility of the system, no ASICs will be developed.  
Instead, a FPGA will be employed to process the data from the two ADCs per 
board and a micro controller to control the FPGA and to allow for the
uploading of upgrades of the FPGA code. 

The acoustic data will be provided with a time stamp derived from the
standard ANTARES clock in order to allow the correlation of the data
from several storeys.
The electric power consumption of the design will be below $8.5$ W per
storey, which is the power available for acoustics per electronics
container.

\section{Outlook and summary}
It is foreseen to install two acoustic sectors with up to 60 acoustic
sensors in total into the ANTARES detector.
For this goal, a conclusive concept has been devised by the Erlangen
group.  It is currently under discussion inside the ANTARES
collaboration where to place the two acoustic sectors.  In the
meantime, the development of ``acoustic ADC-boards'' in Erlangen is
progressing.

\end{document}